\documentclass[12pt,a4paper,english,amssymb,nofootinbib,superscriptaddress]{revtex4}
\usepackage{lmodern}

\usepackage[T1]{fontenc}
\usepackage[latin9]{inputenc}
\usepackage{slashed}
\usepackage{babel}
\usepackage{amsmath}
\usepackage{slashed}
\usepackage{amssymb}
\usepackage{graphicx}
\usepackage{epstopdf}
\usepackage{subfigure}
\usepackage{esint}
\usepackage[unicode=true,pdfusetitle,
bookmarks=true,bookmarksnumbered=false,bookmarksopen=false,
breaklinks=false,pdfborder={0 0 1},backref=false,colorlinks=false]
{hyperref}

\makeatletter
\@ifundefined{textcolor}{}
{%
	\definecolor{BLACK}{gray}{0}
	\definecolor{WHITE}{gray}{1}
	\definecolor{RED}{rgb}{1,0,0}
	\definecolor{GREEN}{rgb}{0,1,0}
	\definecolor{BLUE}{rgb}{0,0,1}
	\definecolor{CYAN}{cmyk}{1,0,0,0}
	\definecolor{MAGENTA}{cmyk}{0,1,0,0}
	\definecolor{YELLOW}{cmyk}{0,0,1,0} 
}
 
\usepackage{latexsym}\usepackage{bm}

\makeatother

\makeatother

\begin{document}
	
\title{A Note on Noncompact and Nonmetricit Quadratic Curvature Gravity Theories}                  
 
\author{Suat Dengiz$^*$ \\
$^*$Correspondence: dengizsuat@gmail.com}     

\date{\today}
 
\begin{abstract}                                            
\textbf{Abstract:} In this note, we evaluate the Weyl-invariant quadratic curvature tensors for the \emph{particular} Weyl's gauge field constructed in the $3+1$-dimensional noncompact Weyl-Einstein-Yang-Mills model. We subsequently extend the model to its higher curvature version. Here, we also compute Weyl-invariant extension of topological Gauss-Bonnet term for this specific choice of vector field. \\

\end{abstract}
\maketitle 
\section{Introduction}                              
  
Unlike its astonishing successes in the intermediate scales, Einstein's general theory of relativity loses its predictability in the IR and UV regimes of universe. More precisely, it is well-known that Einstein's gravity breaks down to explain the accelerated expansion of universe and rotation curves of spiral galaxies in the IR scales. These problems can be fixed by the assumption of substantial amount of dark energy and matter even if they are completely unknown. As for the UV regimes, it has been shown that although pure Einstein's gravity is renormalizable at one loop level, it turns out to be non-renormalizable as one takes all the possible internal matter (that is, scalar, spinor \emph{etc.}) loops into account during the perturbative study \cite{tHooftVeltman, DeserNieuw}. The ensuing computations have demonstrated that the S matrix of bare Einstein's gravity includes uncontrollable UV divergences at two loop-level, too \cite{twoloop1, twoloop2, twoloop3}. So, the theory is an effective one. On the other side, it has been proven that Einstein's gravity augmented with quadratic curvature terms $R^2$ and $R^2_{\mu\nu}$ is renormalizable \cite{Stelle}\footnote{In fact, there is also $R^2_{\mu\nu\alpha\beta}$ term which generally needs to be taken into account. However, since the variation of the topological Gauss-Bonnet term vanishes in $D=3+1$
\begin{equation}
\delta \int d^4 x \sqrt{-\mbox{g}} (R^2-4R^2_{\mu\nu}+R^2_{\mu\nu\alpha\beta})=0,
\label{firsgssbnt}
\end{equation}
one can always eliminate $R^2_{\mu\nu\alpha\beta}$ and only work with the terms $R^2$ and $R^2_{\mu\nu}$.}. In this case, the theory propagates with extra massive spin-0 and spin-2 particles in addition to the massless spin-2 mode belonging to the pure theory. However, the unitarity of massless and massive spin-2 modes here unexpectedly turn out to be in conflict and so the theory becomes non-unitary. This makes the model untrustworthy in quantum theory aspect. [Despite these shortcomings, the higher curvature modifications leave many features of classical Einstein's gravity intact. For instance, as in the Arnowitt-Deser-Misner (ADM) conserved energy and angular momentum \cite{ADM1, ADM2}, one is able to evaluate the conserved quantities dubbed as the Abbott-Deser-Tekin (ADT) charge and superpotential via Killing vectors \cite{ADT1,ADT2,ADT3}\footnote{Here, see also \cite{Tahsin1, Tahsin2} for the intriguing studies concerning $D$-dimensional higher curvature modifications of cosmological Einstein's gravity.}.] At that point, one might stop searching for a complete perturbative quantum gravity theory and only look for a nonperturbative one. One could alternatively move to the beginning and release the ruled out degrees of freedom (DOF) associated to the torsion and nonmetricity in the connection \cite{Eddington, Schrodinger1, Schrodinger2, Hehl}. In this perception, since it ends up with the Weyl's geometry which is a natural place for the local scale-invariant field theories, the specific boost of the nonmetricity seems to be more luring. Recall that the Weyl's method suggests to change the rigid scale invariance that imposes the conformal flatness for the sake of Lorentz symmetry to a \emph{local} scale-invariance to get the Poincare-invariant models in generic spacetimes \cite{Weyl1918, Weyl1919, London, oraif, Iorio}. Due to the fact that the local scale-invariance does not tolerate dimensionful quantities which also turn out to be worthless in the sufficiently high energies according to special relativity, this symmetry might be a fundamental symmetry of universe so that its breaking would generate the dimensionful quantities such as Newton's constant that is the main reason of non-renormalizability in Einstein's gravity. See \cite{Jimenez1, Tanhayindim, Oliva, KilicarslanTMG, TanhayiUnNMG, Jackiw1, Jackiw2, Jackiw3, Grumiller, DengizPhd, KilicarslanPhD} for some interesting works concerning the Weyl's conformal symmetry in various field theories. For the quantization of definite Weyl-invariant gravity models via one-loop beta functions, see \cite{Shapiro, Percacci}.

In \cite{Dengizanphys}, by supposing the Higgs-like field to be the main source of the transition from Einstein's geometry to Weyl's geometry, it has recently been built a unitarity semiclassical quantum gravity model called the $3+1$-dimensional noncompact Weyl-Einstein-Yang-Mills model. Here, the occurring Higgs-like field is defined in $SU(N)$ in the adjoint representation as in the Georgi-Glashow model \cite{Glashw}. With the construction of the Weyl's gauge field from a \emph{particular} superposition of magnitudes of the existing fields in this representation, the Higgs-like sector then supplies the local conformal-invariance to the whole system. Moreover, it is shown that the Weyl's symmetry is spontaneously broken in de Sitter space as in the Higgs mechanism \cite{Higgs, Engler} and radiatively broken at one loop level in flat space \emph{\`a la} Coleman-Weinberg mechanism \cite{ColemanWeinberg}. On the other hand, the model is not unitary in anti-de Sitter space. 

In the current work, since the Weyl-invariant higher order curvature tensors in \cite{DengizTekin} contain extra vector fields and thus provide more information as compared to the bare ones, we will go further and evaluate these higher derivative terms for the \emph{particular} Weyl's gauge field constructed in \cite{Dengizanphys} and accordingly extend the corresponding noncompact model to its higher curvature version. Additionally, we will also evaluate the Weyl-invariant Gauss-Bonnet combination for this particular Weyl's gauge field.

\section{Weyl's Symmetry}     
Since it is one of the corner stones in the construction of the noncompact model in \cite{Dengizanphys}, we shortly provide some fundamentals of the local Weyl's gauging in this part. For this purpose, let us recall that not long after Einstein's establishment of general relativity, Herman Weyl readdressed the theory and lifted the metric-compatibility constraint on the connection with the help of compensating vector potentials $E_\mu$. In this case, the ordinary Levi-Civita connection is upgraded to the following nonmetricit one \cite{Weyl1918, Weyl1919, London, oraif, Iorio}\footnote{In this section, we follow \cite{DengizTekin,Dengizanphys}. For a thorough study of Weyl's approach, see also \cite{Jimenez1,Tanhayindim, DengizPhd}.}
\begin{equation}
	\tilde{\Gamma}^\lambda_{\mu\nu}=\frac{1}{2}g^{\lambda\sigma} ( \tilde{D}_\mu g_{\sigma\nu}+\tilde{D}_\nu g_{\mu\sigma}-\tilde{D}_\sigma g_{\mu\nu} ),
\label{weychrf}
\end{equation}
where the gauge covariant derivative is defined as $ \tilde{D}_\mu g_{\alpha \beta} =\partial_\mu g_{\alpha\beta} + 2 E_\mu g_{\alpha \beta}$. Note that Eq.(\ref{weychrf}) is invariant under the rescaling 
\begin{equation}
g_{\mu\nu} \rightarrow g^{'}_{\mu\nu}=e^{2 \sigma(x)} g_{\mu\nu}, 
\label{weytrnsfddd}
\end{equation}
 throughout which  $E_\mu \rightarrow E^{'}_\mu = E_\mu - \partial_\mu \sigma(x)$. Here, $\sigma(x)$ is a free local function. With this setting, the Weyl-invariant Riemann tensor then reads
\begin{equation}
\begin{aligned}
\tilde{R}^\mu{_{\nu\rho\sigma}} [g,E]&=\partial_\rho \tilde{\Gamma}^\mu_{\nu\sigma}-\partial_\sigma \tilde{\Gamma}^\mu_{\nu\rho}+ \tilde{\Gamma}^\mu_{\lambda\rho} \tilde{\Gamma}^\lambda_{\nu\sigma}-\tilde{\Gamma}^\mu_{\lambda\sigma} \tilde{\Gamma}^\lambda_{\nu\rho} \\
& =R^\mu{_{\nu\rho\sigma}}+\delta^\mu{_\nu}F_{\rho\sigma}+2 \delta^\mu{_[\sigma} \nabla_{\rho]} E_\nu+2 g_{\nu[\rho}\nabla_{\sigma]} E^\mu \\
& \hskip 3.2 cm+2 E_[\sigma   \delta_{\rho]}\,^\mu E_\nu + 2 g_{\nu[\sigma}  E_{\rho]} E^\mu  +2 g_{\nu[\rho} \delta_{\sigma]}\,^\mu  E^2 ,
\label{weinvriem}
\end{aligned} 
\end{equation}
where $ 2 E_{[ \rho} E_{\sigma]} \equiv E_\rho E_\sigma -  E_\sigma E_\rho$ and $E^2= E_\mu E^\mu$. Thereupon, the Weyl-invariant Ricci tensor becomes
\begin{equation}
\begin{aligned} 
\tilde{R}_{\nu\sigma} [g,E]&= \tilde{R}^\mu{_{\nu\mu\sigma}}[g,E]=R_{\nu\sigma}+F_{\nu\sigma}-(n-2)\Big (\nabla_\sigma E_\nu - E_\nu E_\sigma +E^2  g_{\nu\sigma} \Big)-  g_{\nu\sigma} \nabla_\mu  E^\mu.
\end{aligned}
\end{equation}
 Finally, the Weyl-extended curvature scalar comes to be
\begin{equation}
\tilde{R}[g,E]=R-2(n-1)\nabla_\mu  E^\mu-(n-1)(n-2) E^2,
\label{weyltransricciscalar}
\end{equation}
that is \emph{not} Weyl-invariant but instead changes as $\tilde{R}[g, E] \rightarrow (\tilde{R}[g,E])^{'} = e^{-2 \sigma (x) }  \tilde{R}[g, E]$. Then, with the help of a suitably adjusted scalar field, one will obtain the Weyl-invariant Einstein's gravity as follows \cite{DengizTekin,Dengizanphys}
\begin{equation}
\begin{aligned}
S&= \int d^n x \sqrt{-g} \, \Phi^2 \, \tilde{R}[g, E] \\
&= \int d^n x \sqrt{-g} \, \Phi^2 \, \Big [R-2(n-1)\nabla \cdot E-(n-1)(n-2) E^2 \Big].
\label{eh}
\end{aligned}
\end{equation}

As to the lower spin bosonic fields, by suggesting \cite{oraif, Iorio, DengizTekin, Dengizanphys} for details, let us notice that the Weyl-invariant extension of scalar and Maxwell-type field theories respectively are
\begin{equation}
\begin{aligned}
S_\Phi&=- \frac{1}{2}\int d^n x \sqrt{-g}\Big ((\partial_\mu \Phi -\frac{n-2}{2} E_\mu \Phi )^2+\nu \, \Phi^{\frac{2n}{n-2}}\Big ), \\ 
S_{E_\mu} &=  \beta \int d^n x \sqrt{-g}\,\, \Phi^{\frac{2(n-4)}{n-2}} F_{\mu \nu} F^{\mu \nu},
\label{scalarwithpot}
\end{aligned}
\end{equation}
which are invariant under the rescaling in Eq.(\ref{weytrnsfddd}) in addition to $\Phi \rightarrow \Phi^{'} =e^{-\frac{(n-2)}{2}\sigma(x)} \Phi.$ Observe that the generic Weyl-invariant scalar field potential which is renormalizable at least in $2+1$ and $3+1$ dimensions is also assumed in order to recover the cosmological Einstein's gravity after the conformal symmetry is broken. 
   
\section{Higher Curvature Modification of the Noncompact Weyl-Einstein-Yang-Mills Model} 
In this section, we will focus on the quadratic curvature modification of the $3+1$-dimensional Noncompact Weyl-Einstein-Yang-Mills Model in \cite{Dengizanphys}. Recall that here a semiclassical unitary noncompact quantum gravity model in which the Weyl's symmetry of entire system is generated by the existing Higgs-like sector is constructed. To this end, the existing Higgs-like field $\varphi$ is initially supposed to be an element of $SU(N)$ in the \emph{adjoint representation} and subsequently its magnitudes in this generator bases are imposed to act as the Weyl's scalar field:
\begin{equation}
\varphi^a \rightarrow \varphi^{a'}= e^{-\sigma(x)} \varphi^a,
\label{weylcompsca}
\end{equation}
with which the group transformation turns into
\begin{equation}
U \rightarrow {\cal U}=U \,\, e^{-\sigma(x)}.
\label{defgroptranfsg}
\end{equation}
In this case, the Higgs-like field transforms as follows 
\begin{equation}
\varphi \rightarrow \varphi^{'}={\cal U} \varphi {\cal U}^{-1}.
\label{noncmpscfldtrnsf}
\end{equation}
Referring \cite{Dengizanphys} for details, let us note that with this distortion, the compact $SU(N)$ gauge group is being promoted to the noncompact $SL(N, \mathbb{C})$ comprising $2N^2-2$ generators $\{ T^a \cdots, \mbox{i} T^a \cdots \} \,\, \mbox{where} \,\, a=1,2,...,N^2-1$ \cite{Fuchs, Barut, Hsu}. Accordingly, the corresponding noncompact gauge field $A_\mu$ can be expressed in terms of a non-abelian gauge field $B_\mu$ and a gauge covariant field $C_\mu$\footnote{Recall that $C_\mu$ transforms as $C_\mu \rightarrow C^{'}_\mu={\cal U} C_\mu {\cal U}^{-1}$ \cite{Nair}.} as follows
\begin{equation}
 A_\mu \equiv B_\mu+\mbox{i} C_\mu.
 \label{decnongfld}
\end{equation}
Besides, to have the local $SL(N, \mathbb{C})$ invariant scalar field theory, one has to replace the ordinary partial derivative with the following noncompact gauge covariant derivative in the adjoint representation 
\begin{equation}
{\cal D}_\mu \varphi \equiv \partial_\mu \varphi- \mbox{i} g [ A_\mu, \varphi].
 \label{noncompgagcovder}
 \end{equation}
Here, the gauge field transforms as follows
\begin{equation}
A^\mu \rightarrow A^{'}_\mu= {\cal U} A_\mu {\cal U}^{-1}+\frac{1}{\mbox{i} g} (\partial_\mu {\cal U}){\cal U}^{-1}.
\label{nongagtrnfsm}
\end{equation}
As for having a dynamical noncompact gauge field, let us first notice that by considering the ensuing field-strength tensor 
\begin{equation}
 F_{\mu\nu}=\partial_\mu A_\nu-\partial_\nu A_\mu- \mbox{i}g[A_\mu, A_\nu] \hskip 0.5 cm \mbox{with} \hskip 0.5 cm F_{\mu\nu} \rightarrow F^{'}_{\mu\nu}={\cal U} F_{\mu\nu} {\cal U}^{-1},
 \label{kinnoncmp}
\end{equation}
one will be able to expand Eq.(\ref{kinnoncmp}) in the generator bases of $SU(N)$ by virtue of $ A_\mu \equiv A^a_\mu T^a =(B^a_\mu+ \mbox{i} C^a_\mu)T^a$ as  follows
 \begin{equation}
F_{\mu\nu}=F_{\mu\nu}^a T^a  \hskip 0.5 cm \mbox{where} \hskip 0.5 cm F^a_{\mu\nu}={\cal B}^a_{\mu\nu}+\mbox{i} {\cal C}^a_{\mu\nu},
 \label{expnonfstts} 
\end{equation}
in which one has
\begin{equation}
 \begin{aligned}
{\cal B}^a_{\mu\nu}&= \partial_\mu B^a_\nu-\partial_\nu B^a_\mu+g f^{abc} (B^b_\mu B^c_\nu-C^b_\mu C^c_\nu), \\
 {\cal C}^a_{\mu\nu}&= \partial_\mu C^a_\nu-\partial_\nu C^a_\mu+g f^{abc} (B^b_\mu C^c_\nu+C^b_\mu B^c_\nu).
 \label{fstnoncompcast3} 
 \end{aligned}
\end{equation}
Then, by using these settings as well as the Higgs-like-dependent function $\Theta=\Theta \,[\varphi^a(x)] $ which is defined during the redefinition of the gamma matrices
\begin{equation}
\Gamma_\mu(x)=\gamma^\mu \Theta \,[\varphi^a(x)],
\end{equation}
in order to get the Weyl-Yang-Mills invariant Dirac theory\footnote{See Appendix for the details of the construction of the Weyl-Yang-Mills invariant Dirac theory.}, one will finally get the noncompact kinetic term as follows \cite{Dengizanphys, Hsu} 
\begin{equation}
\begin{aligned}
\mbox{Tr} (F^{+}_{\mu\nu} \Theta F^{\mu\nu} \Theta^{-1} )&=-2 F^{+a}_{\mu\nu} F^{a\mu\nu}+\frac{4 \Theta^a \Theta^b}{{\bf \Theta}^2} F^{+a}_{\mu\nu} F^{b\mu\nu} \\
&= -2 \Big ({\cal B}^a_{\mu\nu}{\cal B}^{a\mu\nu}+{\cal C}^a_{\mu\nu}{\cal C}^{a\mu\nu}\Big)+\frac{4 \Theta^a \Theta^b}{{\bf \Theta}^2} \Big ({\cal B}^a_{\mu\nu}{\cal B}^{b\mu\nu}+{\cal C}^a_{\mu\nu}{\cal C}^{b\mu\nu}\Big).
\label{kintermforgagttttt}
\end{aligned}
\end{equation}

Observe that all the settings so far are not adequate to generate the local Weyl's symmetry from the Higgs-like sector unless an appropriate Weyl's gauge field is also constructed from these ingredients. As is demonstrated in \cite{Dengizanphys}, let us recall that by defining the following specific combination of the magnitudes of Higgs-like and gauge covariant fields in the generator bases (that is, $\varphi^a$ and $C^a$, respectively) as the Weyl's gauge field 
 \begin{equation}
	E_\mu=g f^{abc} C^a_\mu \varphi^b (\varphi^c)^{-1},   
	\label{specificweylgfldl}
\end{equation}
the Higgs-like field will eventually supply the Weyl's conformal symmetry to the gravity sector consistently. Notice that all the generator indices are entirely closed and since all the ingredients are real variables, then, Eq.(\ref{specificweylgfldl}) automatically becomes real. Observe that $\sigma(x)$ in Sec.II must be chosen as follows
\begin{equation}
\sigma(x)=- g f^{abc} f^{klm} \int dx^{\mu} C^a_\mu (x) \varphi^b (x) (\varphi^c)^{-1}(x) \, w^k(x)w^l(x)T^m+{\cal H},  
\end{equation}	
where ${\cal H}$ is any free constant. Note that with Eq.(\ref{specificweylgfldl}), the Weyl-extended curvature scalar in Eq.(\ref{weyltransricciscalar}) turns into
\begin{equation}
\begin{aligned}
	\tilde{R}[g_{\mu\nu}, E_\mu] &=R-6 g f^{abc} \nabla_\mu C^{a\mu} \varphi^b (\varphi^c)^{-1} \\
	&\hskip 0.82 cm -6 g^2 f^{abc} f^{klm} C^a_\mu \varphi^b (\varphi^c)^{-1} \, C^{k\mu} \varphi^l (\varphi^m)^{-1},
	\label{manfeq}
\end{aligned}
\end{equation}
and the gauge covariant derivatives of metric and scalar fields respectively become
\begin{equation}
\tilde{D}_\mu g_{\alpha\beta}=\partial_\mu g_{\alpha\beta}+2g f^{abc} C^a_\mu \varphi^b (\varphi^c)^{-1} g_{\alpha\beta}, \hskip 0.7 cm 
\tilde{D}_\mu \Phi=\partial_\mu \Phi-g f^{abc} C^a_\mu \varphi^b (\varphi^c)^{-1} \Phi.
\end{equation}
At this point, let us go further by noticing that, with the definite Weyl's gauge field in Eq.(\ref{specificweylgfldl}), the Weyl-invariant quadratic curvature tensors in \cite{DengizTekin} respectively read as follows: Firstly, the Weyl-invariant curvature scalar square becomes
\begin{equation}
\begin{aligned}
\tilde{R}^2[g_{\mu\nu}, E_\mu]&= R^2-12 g R f^{abc}(\nabla^\mu  C^a_\mu) \varphi^b (\varphi^c)^{-1}-12 g^2R f^{abc} f^{klm} C^a_\mu \varphi^b (\varphi^c)^{-1} \, C^{k\mu} \varphi^l (\varphi^m)^{-1}\\
&+36 g^2  f^{abc}f^{klm}(\nabla^\mu  C^a_\mu) \varphi^b (\varphi^c)^{-1}(\nabla^\nu  C^k_\nu) \varphi^l (\varphi^m)^{-1}\\
& +72 g^3f^{abc} f^{klm} f^{npr} C^a_\mu \varphi^b (\varphi^c)^{-1} \, C^{k\mu} \varphi^l (\varphi^m)^{-1} (\nabla^\nu  C^n_\nu) \varphi^p (\varphi^r)^{-1} \\
&+36 g^4 f^{abc} f^{klm} f^{npr} f^{stu} C^a_\mu \varphi^b (\varphi^c)^{-1} \, C^{k\mu} \varphi^l (\varphi^m)^{-1} C^n_\nu \varphi^p (\varphi^r)^{-1} \, C^{s\nu} \varphi^t (\varphi^u)^{-1}.
\label{weylsqrcrv1}
\end{aligned}
\end{equation}
Secondly, the Weyl-invariant Ricci square turns into
\begin{equation}
\begin{aligned}
\tilde{R}^2_{\mu\nu}[g_{\mu\nu}, E_\mu]&= R^2_{\mu\nu}-4 g f^{abc} R^{\mu\nu} (\nabla_\mu  C^a_\nu) \varphi^b (\varphi^c)^{-1} -2g R f^{abc}(\nabla^\mu  C^a_\mu) \varphi^b (\varphi^c)^{-1}\\
&+4 g^2  f^{abc} f^{klm} R^{\mu\nu} C^a_\mu \varphi^b (\varphi^c)^{-1} \, C^k_\nu \varphi^l (\varphi^m)^{-1} \\
&-4 g^2 R f^{abc} f^{klm} C^a_\mu \varphi^b (\varphi^c)^{-1} \, C^{k\mu} \varphi^l (\varphi^m)^{-1}\\
&+F_{\mu\nu}^2-4 gf^{abc} F^{\mu\nu} (\nabla_\nu  C^a_\mu) \varphi^b (\varphi^c)^{-1} \\
& +2 g^2 f^{abc} f^{klm} (\nabla_\nu  C^a_\mu) \varphi^b (\varphi^c)^{-1}(\nabla^\nu  C^{k\mu}) \varphi^l (\varphi^m)^{-1}\\
&+8 g^2 f^{abc} f^{klm} (\nabla^\mu  C^a_\mu) \varphi^b (\varphi^c)^{-1}(\nabla^\nu  C^k_\nu) \varphi^l (\varphi^m)^{-1}\\
&-8 g^3  f^{abc} f^{klm}f^{npr} C^a_\mu \varphi^b (\varphi^c)^{-1} \, C^k_\nu \varphi^l (\varphi^m)^{-1} (\nabla^\mu  C^{n\nu}) \varphi^p (\varphi^r)^{-1} \\
&+20 g^3 f^{abc} f^{klm} f^{npr} C^a_\mu \varphi^b (\varphi^c)^{-1} \, C^{k\mu} \varphi^l (\varphi^m)^{-1} (\nabla^\nu C^n_\nu) \varphi^p (\varphi^r)^{-1} \\
&+12 g^4 f^{abc} f^{klm} f^{npr} f^{stu} C^a_\mu \varphi^b (\varphi^c)^{-1} \, C^{k\mu} \varphi^l (\varphi^m)^{-1} C^n_\nu \varphi^p (\varphi^r)^{-1} \, C^{s\nu} \varphi^t (\varphi^u)^{-1}.
\label{weylsqrcrv2}
\end{aligned}
\end{equation}
Finally, the Weyl-gauged Riemann square becomes
\begin{equation}
\begin{aligned}
\tilde{R}^2_{\mu\nu\rho\sigma}[g_{\mu\nu}, E_\mu] &=R^2_{\mu\nu\rho\sigma}-8g  f^{abc}  R^{\mu \nu} (\nabla_\mu  C^a_\nu) \varphi^b (\varphi^c)^{-1} \\
&+8 g^2 f^{abc} f^{klm} R^{\mu \nu}  C^a_\mu \varphi^b (\varphi^c)^{-1} \, C^k_\nu \varphi^l (\varphi^m)^{-1}\\
&-4g^2 R f^{abc} f^{klm} C^a_\mu \varphi^b (\varphi^c)^{-1} \, C^{k\mu} \varphi^l (\varphi^m)^{-1}\\
& + 4 F_{\mu \nu}^2+8 g^2 f^{abc} f^{klm} (\nabla_\mu  C^a_\nu) \varphi^b (\varphi^c)^{-1}(\nabla^\mu  C^{k\nu}) \varphi^l (\varphi^m)^{-1} \\
&+4g^2 f^{abc} f^{klm} (\nabla^\mu  C^a_\mu) \varphi^b (\varphi^c)^{-1}(\nabla^\nu  C^k_\nu) \varphi^l (\varphi^m)^{-1}\\
&+16 g^3 f^{abc} f^{klm} f^{npr} C^a_\mu \varphi^b (\varphi^c)^{-1} \, C^{k\mu} \varphi^l (\varphi^m)^{-1} (\nabla^\nu C^n_\nu) \varphi^p (\varphi^r)^{-1}\\
&-16 g^3  f^{abc} f^{klm}f^{npr} C^a_\mu \varphi^b (\varphi^c)^{-1} \, C^k_\nu \varphi^l (\varphi^m)^{-1} (\nabla^\mu  C^{n\nu}) \varphi^p (\varphi^r)^{-1}\\
&+12g^4 f^{abc} f^{klm} f^{npr} f^{stu} C^a_\mu \varphi^b (\varphi^c)^{-1} \, C^{k\mu} \varphi^l (\varphi^m)^{-1} C^n_\nu \varphi^p (\varphi^r)^{-1} \, C^{s\nu} \varphi^t (\varphi^u)^{-1}.
\label{weylsqrcrv3}
\end{aligned}
\end{equation}
Notice that with these particular Weyl-extended quadratic curvature tensors, one gets the Weyl-invariant extension of the topological Gauss-Bonnet combination as follows
\begin{equation}
\begin{aligned}
\tilde{R}^2_{\mu\nu\rho\sigma}-4\tilde{R}^2_{\mu\nu}+\tilde{R}^2&=R^2_{\mu\nu\rho\sigma}-4 R^2_{\mu\nu}+R^2+8g f^{abc} R^{\mu\nu}(\nabla_\mu  C^a_\nu) \varphi^b (\varphi^c)^{-1}\\
&-8 g^2 f^{abc} f^{klm} R^{\mu\nu}  C^a_\mu \varphi^b (\varphi^c)^{-1} \, C^k_\nu \varphi^l (\varphi^m)^{-1} \\
& -8 g^2  f^{abc} f^{klm} (\nabla_\mu  C^a_\nu) \varphi^b (\varphi^c)^{-1}(\nabla^\mu  C^{k\nu}) \varphi^l (\varphi^m)^{-1} \\
&  +8g^2f^{abc} f^{klm} (\nabla^\mu  C^a_\mu) \varphi^b (\varphi^c)^{-1}(\nabla^\nu  C^k_\nu) \varphi^l (\varphi^m)^{-1} \\
&+8g^3f^{abc} f^{klm} f^{npr} C^a_\mu \varphi^b (\varphi^c)^{-1} \, C^{k\mu} \varphi^l (\varphi^m)^{-1} (\nabla^\nu C^n_\nu) \varphi^p (\varphi^r)^{-1}\\
&  + 16g^3 f^{abc} f^{klm}f^{npr} C^a_\mu \varphi^b (\varphi^c)^{-1} \, C^k_\nu \varphi^l (\varphi^m)^{-1} (\nabla^\mu  C^{n\nu}) \varphi^p (\varphi^r)^{-1}\\
&-4g R f^{abc}(\nabla^\mu  C^a_\mu) \varphi^b (\varphi^c)^{-1}.           
\end{aligned}
\label{gb} 
\end{equation}
Observe that all the representation indices in the above-given quantities are entirely contracted.

By referring \cite{Dengizanphys} for further details, let us lastly notice that by taking the magnitudes of Higgs-type field in the adjoint bases  $\varphi^a$ and Eq.(\ref{specificweylgfldl}) as Weyl's scalar and gauge fields, the $3+1$-dimensional noncompact Weyl-Einstein-Yang-Mills in \cite{Dengizanphys} can be modified by the Weyl-invariant quadratic curvature terms in Eq.(\ref{weylsqrcrv1})-(\ref{weylsqrcrv3}) as follows
\begin{equation}
	\begin{aligned}
		{\cal S}_{nWEYM} = \int d^4 x \sqrt{-g}  \bigg \{& \alpha (\varphi^a)^2  \tilde{R}[g_{\mu\nu},E_\mu]+\beta \tilde{R}^2[g_{\mu\nu},E_\mu]+\gamma \tilde{R}^2_{\mu\nu}[g_{\mu\nu},E_\mu]+\rho \tilde{R}^2_{\mu\nu\rho\sigma}[g_{\mu\nu},E_\mu]\\ 
	&+\sigma ({\cal D}_\mu \varphi^a)({\cal D}^\mu \varphi^a)^{+}+ \kappa (\varphi^a)^4+\eta (\varphi^a)^2 \bar{\psi}_b \, \mbox{i} (\slashed{\cal D} \psi )^b \\
&+\zeta \Big[-2 \Big ({\cal B}^a_{\mu\nu}{\cal B}^{a\mu\nu}+{\cal C}^a_{\mu\nu}{\cal C}^{a\mu\nu}\Big)+\frac{4 \Theta^a \Theta^b}{{\bf \Theta}^2} \Big ({\cal B}^a_{\mu\nu}{\cal B}^{b\mu\nu}+{\cal C}^a_{\mu\nu}{\cal C}^{b\mu\nu}\Big) \Big]\bigg \},
		\label{maineqn}
	\end{aligned}
\end{equation}
where $\alpha, \beta, \gamma, \rho, \sigma, \kappa, \eta$ and $\zeta$ are arbitrary dimensionless couplings as expected. Here, the noncompact gauge covariant derivative is   
\begin{equation}
{\cal D}_\mu \varphi^a =\partial_\mu \varphi^a +g f^{abc} A^b_\mu \varphi^c=\partial_\mu \varphi^a +g f^{abc} B^b_\mu \varphi^c+\mbox{i} g f^{abc} C^b_\mu \varphi^c.
\end{equation}
Observe that unlike the generic $D-$dimensional Weyl-invariant higher derivative gravity theories \cite{Tanhayindim, DengizTekin}, here the quadratic curvature terms do not bring any extra scalar field. Now that we have built a legitimate higher derivative extension of the noncompact model, we need to go further and particularly study its unitarity to see if Eq.(\ref{maineqn}) is a viable model at least at the semiclassical level as in \cite{Dengizanphys}. However, as is seen above, since the Weyl-gauged quadratic curvature terms in Eq.(\ref{maineqn}) together with the tools in the bare model are extremely complicated, we restrict ourselves to the construction of the corresponding action in this note and thus leave that task to a separate study.

In summary, in this study, we have computed the Weyl-gauged higher order curvature terms for the specific Weyl's gauge field in the recent noncompact gravity in \cite{Dengizanphys}. By using these tools, we have promoted the noncompact model to its higher order curvature extension. In addition to these, we have also evaluated the Weyl-gauged Gauss-Bonnet combination for this definite Weyl's gauge field.

\section{Acknowledgments}
We would like to thank Bayram Tekin, Tahsin Cagri Sisman and Ercan Kilicarslan for useful suggestions.

\section{Appendix: The Construction of Weyl-Yang-Mills invariant noncompact Kinetic Term for Gauge Field}
In this section, we review the construction of the Weyl-Yang-Mills (WYM) invariant kinetic term for the noncompact gauge field which is shown to provide a unitarity model at least at the tree-level \cite{Dengizanphys}. To do so, let us recall that since the ordinary Yang-Mills-type kinetic term would violate the unitarity, one cannot assume such a term. Subsequently, by observing Eq.(\ref{expnonfstts}), one will see that a kinetic term with following structure seems to be compatible with the unitarity  
\begin{equation}
	\begin{aligned}
	 {\cal L}_{A_\mu} \sim \mbox{Tr}(F_{\mu\nu}F^{+\mu\nu}) \sim {\cal B}^a_{\mu\nu}{\cal B}^{a\mu\nu}+{\cal C}^a_{\mu\nu}{\cal C}^{a\mu\nu}.
 \label{kintermforgag}
	\end{aligned}
\end{equation}
But, Eq.(\ref{kintermforgag}) is not WYM invariant at this time. Since it is partially problematic, an appropriate modification of Eq.(\ref{kintermforgag}) can provide the desired kinetic term. As is shown in \cite{Dengizanphys, Hsu}, this can be achieved with the help of Dirac field as follows: let us first observe that the Dirac theory 
\begin{equation}
{\cal L}_{Dirac}=\bar{\psi} \mbox{i} \gamma^\mu {\cal D}_\mu \psi \quad \mbox{where} \quad {\cal D} =\partial_\mu \psi-\mbox{i} g A_\mu \psi, 
\end{equation}
fails to be invariant under WYM transformations. Here, recall that the Dirac field and its gauge covariant derivative transform as follows 
\begin{equation}
\psi \rightarrow \psi^{'}= {\cal U} \psi, \hskip 1 cm  ( {\cal D}_\mu \psi ) \rightarrow ( {\cal D}_\mu \psi )^{'} = {\cal U} ({\cal D}_\mu \psi ).
\end{equation}
Now, in order to reach our ultimate aim, one needs to replace the ordinary gamma matrices with the following one
\begin{equation}
\gamma_\mu(x) \rightarrow \Gamma_\mu (x) \qquad \mbox{where} \qquad \Gamma_\mu(x) \rightarrow \Gamma^{'}_\mu(x) =({\cal U}^{+})^{-1}\Gamma_\mu(x) {\cal U}^{-1}, 
\end{equation}
that demands a compensating field. Hence, by taking this additional DOF to be a function of the magnitudes of Higgs-like field in the adjoint representation in \cite{Dengizanphys} as 
\begin{equation}
\Gamma_\mu(x)=\gamma^\mu \Theta \,[\varphi^a(x)],
\end{equation}
one will get the WYM-invariant Dirac theory as follows
\begin{equation}
S_{Dirac} = \eta \int d^4 x \sqrt{-g} \, (\varphi^a)^2 \bar{\psi} i \Gamma^\mu {\cal D}_\mu \psi.
\end{equation}
From these tools, one then can define the following noncompact WYM-invariant kinetic term \cite{Dengizanphys, Hsu}
\begin{equation}
\begin{aligned}
\mbox{Tr} (F^{+}_{\mu\nu} \Theta F^{\mu\nu} \Theta^{-1} )&=-2 F^{+a}_{\mu\nu} F^{a\mu\nu}+\frac{4 \Theta^a \Theta^b}{{\bf \Theta}^2} F^{+a}_{\mu\nu} F^{b\mu\nu} \\
&= -2 \Big ({\cal B}^a_{\mu\nu}{\cal B}^{a\mu\nu}+{\cal C}^a_{\mu\nu}{\cal C}^{a\mu\nu}\Big)+\frac{4 \Theta^a \Theta^b}{{\bf \Theta}^2} \Big ({\cal B}^a_{\mu\nu}{\cal B}^{b\mu\nu}+{\cal C}^a_{\mu\nu}{\cal C}^{b\mu\nu}\Big).
\label{kintermforgagttttt}
\end{aligned}
\end{equation}

\end{document}